\documentclass[aps,prc,floatfix,nofootinbib,preprintnumbers,superscriptaddress,longbibliography,notitlepage,twocolumn]{revtex4-2}
\usepackage[dvipsnames,svgnames,x11names]{xcolor}
\usepackage{epsfig}
\usepackage{graphicx}
\usepackage{stmaryrd}
\usepackage{amssymb,tabularx,dcolumn}
\usepackage{supertabular,ltxtable}
\usepackage{longtable}
\usepackage{amsmath}
\usepackage{amstext}
\usepackage{float}
\usepackage{color}
\usepackage{bm}
\usepackage{hyperref}
\hypersetup{breaklinks=true,colorlinks=true,linkcolor=blue,citecolor=blue,filecolor=magenta,urlcolor=cyan}
\usepackage{mathtools}
\usepackage{multirow}
\usepackage{makecell}
\usepackage[utf8]{inputenc}
\usepackage{tabu}
\usepackage{braket}
\usepackage{subfigure} 
\usepackage{enumitem}
\usepackage[english]{babel}

\definecolor{pastelgray}{rgb}{0.81, 0.81, 0.77}
\definecolor{beaublue}{rgb}{0.9, 0.9, 0.93}

\begin{document}

\title{Efficient calculation of two-neutrino double-beta-decay nuclear matrix elements}

\author{M. Horoi}
\affiliation{Department of Physics, Central Michigan University, Mount Pleasant, MI 48859, USA}

\date{\today}

\begin{abstract}
Reliable nuclear matrix elements (NMEs) are essential for interpreting double-beta-decay experiments and for connecting measured or constrained half-lives to the underlying weak-interaction physics. The two-neutrino mode ($2\nu\beta\beta$) is allowed by the Standard Model and has been observed in several nuclei, whereas the neutrinoless mode ($0\nu\beta\beta$) remains the key experimental signature of lepton-number violation and Majorana neutrino masses. Recent statistical shell-model studies indicate a strong correlation between the $2\nu\beta\beta$ and $0\nu\beta\beta$ NMEs, making accurate and efficient calculations of the former especially useful for assessing the latter. Direct evaluations of $2\nu\beta\beta$ NMEs usually require summing over many $1^+$ states in the intermediate odd-odd nucleus, a procedure that becomes expensive and may converge slowly in large model spaces. We present and test an improved strength-function method based on Lanczos iterations that avoids full diagonalization while preserving the accuracy of explicit summation where such benchmarks are possible. The method is applied to several experimentally important emitters and to different effective Hamiltonians. We also show that the same framework can be used for the higher-order NMEs entering Taylor-expanded phase-space treatments of $2\nu\beta\beta$ and related decay modes.
\end{abstract}

\maketitle

\section{Introduction} \label{intro}

Double-beta decay (DBD) is one of the most sensitive low-energy probes of the interplay between nuclear structure and fundamental symmetries. The two-neutrino mode ($2\nu\beta\beta$) is a second-order weak process allowed by the Standard Model and observed in several even-even nuclei. The neutrinoless mode ($0\nu\beta\beta$), if observed, would demonstrate lepton-number violation and establish that neutrinos are Majorana particles~\cite{Avignone2008,Vergados2012}. In both cases, the connection between a measured half-life and the underlying physics depends on phase-space factors (PSFs), weak couplings, and nuclear matrix elements (NMEs)~\cite{Doi1983,Doi1985,SuhonenCivitarese1998,Kotila2012, StoicaMirea2013}.

The experimental program has entered a regime in which theoretical uncertainties in the NMEs have become a central limitation. Current and next-generation searches aim to improve sensitivities from the $10^{26}$ yr scale toward $10^{28}$ yr, with the goal of covering the inverted neutrino-mass ordering region for the standard light-neutrino-exchange mechanism~\cite{EngelMenendez2017,whiteppbb-2022}. For $0\nu\beta\beta$, the inverse half-life contains both the NME and lepton-number-violating parameters associated with the assumed mechanism~\cite{Doi1985,Vergados2012,Rodejohann2012,Deppisch2012}. In the absence of a signal, experimental limits are therefore translated into limits on beyond-the-Standard-Model parameters only after adopting calculated NMEs, often under the assumption that a single mechanism dominates~\cite{18ho035502}.

PSFs are now calculated with relatively high precision using realistic relativistic electron wave functions~\cite{Kotila2012, StoicaMirea2013}. The NMEs remain more difficult because they depend on many-body correlations, model-space truncations, effective interactions, and effective weak operators. A broad range of nuclear-structure approaches has been developed, including the interacting shell model~\cite{Caurier1990,Caurier2005, HoroiStoicaBrown2007}, pn-QRPA~\cite{SuhonenCivitarese1998, Simkovic1999}, the interacting boson model~\cite{Barea2009}, energy-density-functional methods~\cite{Rodriguez2010}, and recent ab initio or coupled-cluster calculations~\cite{Yao2020,Novario2021,Belley2021}. For $0\nu\beta\beta$, different methods can still differ substantially for the same isotope~\cite{EngelMenendez2017,Yao2020,Novario2021}. For $2\nu\beta\beta$, the NME is governed by products of Gamow-Teller (GT) transition amplitudes through $1^+$ states in the intermediate odd-odd nucleus, and realistic comparisons with data usually require an effective quenching of the GT operator~\cite{HoroiStoicaBrown2007}.

The shell model is particularly attractive for DBD because it treats correlations among the active valence nucleons explicitly and preserves the quantum numbers needed for detailed spectroscopy. Its main limitation is the rapid growth of the many-body basis. The conventional calculation of $M_{2\nu}$ uses an explicit sum over intermediate $1^+$ states. This approach is transparent and benchmarkable, but in heavy systems or enlarged valence spaces the number of contributing states can be very large and the running sum may converge slowly or non-monotonically; recent large-scale calculations provide examples of this difficulty~\cite{PhysRevC.110.054323}.

This issue has practical importance beyond the $2\nu\beta\beta$ half-life itself. In a recent statistical shell-model analysis, variations of $0\nu\beta\beta$ NMEs were compared with variations of other observables, including $2\nu\beta\beta$ NMEs, GT strengths, and low-lying spectra~\cite{Horoi-prc22}. The analysis showed a very strong correlation between the $2\nu\beta\beta$ and $0\nu\beta\beta$ NMEs. A reliable and efficient calculation of $M_{2\nu}$ can therefore help identify effective Hamiltonians and operator renormalizations that are more favorable for $0\nu\beta\beta$ predictions, especially if a future $0\nu\beta\beta$ signal is observed in more than one isotope.

The purpose of this work is to present an efficient implementation of the strength-function method for $2\nu\beta\beta$ NMEs and to demonstrate its accuracy by comparison with explicit summations where those are feasible. The method follows the formalism of Ref.~\cite{ehv92}, using Lanczos iterations initiated by GT doorway states, and implements the coefficients of the expansion through a simple numerical prescription recently described in Ref.~\cite{physics4040074}. It avoids storing the eigenvectors of the full many-body Hamiltonian and requires only the Lanczos vectors and the much smaller tridiagonal Lanczos matrix. This gives a significant computational advantage over direct diagonalization and improves on earlier strength-function implementations~\cite{Caurier1990}.

We apply the method to several nuclei of experimental interest, including $^{48}$Ca, $^{76}$Ge, $^{82}$Se, $^{128,130}$Te, and $^{136}$Xe, using standard effective Hamiltonians in the corresponding valence spaces. The calculated running sums are compared with direct summations over hundreds or thousands of intermediate states. We also extend the method to the higher-order NMEs $M_{2\nu-3}$ and $M_{2\nu-5}$ that enter Taylor-expanded treatments of PSFs and related two-neutrino electron-capture observables~\cite{Niţescu_2024}. The paper is organized as follows. Section \ref{half-life} summarizes the half-life formalism and the strength-function algorithm. Section \ref{results} presents benchmark calculations and isotope-by-isotope results. Section \ref{Taylor} discusses the Taylor-expansion NMEs. Section \ref{conclusions} gives the conclusions and outlook.

\begin{center}
\begin{table*}[htb]
\begin{tabular}{lcccccccl}
\hline
Isotope & $Q_{\beta\beta}$ & $Q_{\beta}$ & $E_0$ & $E_1(1^+)$ & $M^{eff}_{2\nu}$ & $M_{2\nu}$ & $q_{opt}$ & Hamiltonian\\
\hline
$^{48}$Ca & 4.268 & 0.280 & 1.855 & 2.517 & 0.0350 & 0.0421 & 0.68 & GXPF1A \\
$^{76}$Ge & 2.039 & -0.922 & 1.941 & 0.044 & 0.1060 & 0.1274  & 0.63 & GCN2850 \\
&&&&&&& 0.62 & JUN45 \\
$^{82}$Se & 2.998 & -0.095 & 1.594 & 0.075 & 0.0850 & 0.1022 & 0.55 & GCN2850 \\
&&&&&&& 0.62 & JUN45 \\
$^{96}$Zr & 3.356 & 0.164 & 1.514 &  ? & 0.0800 & 0.0962  & & \\
$^{100}$Mo & 3.034 & -0.172 & 1.689 & 0.000 & 0.1850 & 0.2224  & & \\
$^{116}$Cd & 2.813 & -0.463 & 1.869 & 0.000 & 0.1080 & 0.1298  & & \\
$^{128}$Te & 0.867 & -1.256 & 1.689 & 0.000 & 0.0430 & 0.0517  & 0.84 & SVD \\ 
$^{130}$Te & 2.528 & -0.417 & 1.681 & 0.255 & 0.0293 & 0.0352  & 0.50 & GCN5082 \\
&&&&&&& 0.88 & SVD \\
$^{136}$Xe & 2.458 & -0.090 & 1.319 & 0.590 & 0.0181 & 0.0218 & 0.42 & GCN5082 \\
&&&&&&& 0.69 & SVD \\
$^{150}$Nd & 3.371 & -0.083 & 1.769 &  ? & 0.0550 & 0.0661  & & \\
$^{238}$U & 1.114 & -0.147 & 0.704 & 0.244 & 0.1300 & 0.1563  & & \\

\hline
\end{tabular}
\label{qfactors}
\caption{Input data used for calculating $M_{2\nu}$ and for comparison with the effective matrix elements $M^{eff}_{2\nu}$ recommended in Table 3 of Ref.~\cite{Barabash2020}. All energies are in MeV and $M_{2\nu}$ is in MeV$^{-1}$. The free-nucleon axial weak coupling constant used in converting between $M^{eff}_{2\nu}$ and $M_{2\nu}$ is $g_A=1.275$. The valence spaces and Hamiltonian acronyms are defined in the text.}
\end{table*}
\end{center}

\section{The $2\nu\beta\beta$ decay half-life} \label{half-life}

Within the Standard Model, two-neutrino double beta decay serves as the allowed counterpart to the neutrinoless mode. Having been detected in various isotopes (see Table~\ref{qfactors}), this process offers a critical benchmark for validating the Gamow-Teller (GT) components of nuclear wave functions employed in double beta decay research. This section outlines the half-life formalism adopted here and details a strength-function methodology designed to compute nuclear matrix elements (NMEs) in large-scale model spaces where the traditional approach of summing over all intermediate states is computationally prohibitive.

For a transition originating from an initial $0^+$ ground state and terminating in a final $J=0$ state of the granddaughter nucleus, the GT NME is expressed as~\cite{Tomoda1991}

\begin{equation}
M_{2\nu} = 
\sum_k \frac{\langle 0^+_f\vert q \sigma\tau^-\vert 1^+_{k}
\rangle 
\langle 1^+_{k}\vert q \sigma\tau^-
\vert 0^+_i\rangle}{E_{rel,1_1^+}(1_k^+) +E(1_1^+)+ E_0} .
\label{eq1}
\end{equation}

\noindent 
In this expression, $E_{rel,1_1^+}(1_k^+)$ denotes the excitation energy of the $k$-th $1^+$ state in the intermediate odd-odd nucleus relative to the lowest-lying $1^+$ state, $E(1_1^+)$. Utilizing the first $1^+$ state as an energy anchor is a common practice, as its value is often available from experimental data (refer to Table~\ref{qfactors}). The term $E_0 = \frac{1}{2}Q_{\beta\beta} + \Delta M$ incorporates the $Q$-value for the transition to the final $0^+_f$ state and the mass difference $\Delta M$ between the parent and intermediate nuclei. Alternatively, $E_0$ can be expressed as $\frac{1}{2}Q_{\beta\beta} - Q_{\beta^-}$ by utilizing the beta-minus $Q$-value (refer to Table~\ref{qfactors}). While transitions to the ground state ($0^+_1$) are most frequent, decays to excited $0^+_2$ states have also been documented~\cite{Vergados2012}. The quenching parameter $q$ is used to account for two-body currents, missing correlations, and the specific choice of valence space and Hamiltonian. Based on comparisons with experimental matrix elements for the Hamiltonians used in this study, $q$ typically ranges from $0.40$ to $0.88$, as detailed in Table~\ref{qfactors}.

The inverse of the $2\nu\beta\beta$ half-life is defined as:

\begin{equation}
\left[ T^{2\nu}_{1/2} \right]^{-1} = G_{2\nu} \cdot \left[ g_A^2  \left(m_ec^2 \cdot M_{2\nu}\right) \right]^2 \equiv G_{2\nu} \left( M_{2\nu}^{eff} \right)^2
\label{2nu-hl}
\end{equation}

\noindent
Previous work in Ref.~\cite{HoroiStoicaBrown2007} utilized the full diagonalization of 250 $1^+$ states in $^{48}$Sc (within the $pf$ shell) to determine the NME for $^{48}$Ca. While such direct diagonalization is feasible for certain nuclei using J-scheme codes like NuShellX \cite{nushellx}, it becomes impractical as the model space expands. For instance, including $2\hbar\omega$ excitations in the $sd-pf$ space for $^{48}$Ca leads to m-scheme dimensions exceeding 1 billion (with 716 million for $^{48}$Sc), necessitating a more scalable approach. Although the strength-function method introduced in Ref.~\cite{Caurier1990} for $^{48}$Ca converges rapidly, it still requires exhaustive state calculations to establish rigorous convergence benchmarks. To address this, Ref.~\cite{ehv92} proposed a framework to derive energy-weighted sums from Lanczos strength functions, though it lacked a comprehensive numerical recipe for shell-model implementation. We therefore employ the simplified implementation of this expansion recently detailed in Ref.~\cite{physics4040074}.

Following the logic of Ref.~\cite{ehv92}, we define the starting Lanczos vectors $L^{\pm}_1$ by applying the Gamow-Teller operator to either the initial or final $0^+$ states:

\begin{eqnarray}
\label{stm}
|\sigma \tau^- 0^+_i> = c_- |dw_- > \equiv c_- |L^-_1 > \\
\label{stp}
|\sigma \tau^+ 0^+_f> = c_+ |dw_+ > \equiv c_+ |L^+_1 >\ .
\end{eqnarray}

\noindent
Here, $|dw_{\pm}>$ are the doorway states and $c_{\pm}$ correspond to the square roots of the GT sum rules. According to Ref.~\cite{ehv92}, the NME in Eq.~(\ref{eq1}) can be approximated via either of the following sums over Lanczos vectors $L_m$:

\begin{eqnarray}
\label{mm}
M_{2\nu}(0^+) \approx 3 c_+ c_- \sum_m g^-_m <dw_+|L^-_m> \equiv M_{2\nu}^{GT-}\\
\label{mp}
M_{2\nu}(0^+) \approx 3 c_+ c_- \sum_m g^+_m <L^+_m|dw_-> \equiv M_{2\nu}^{GT+}\ .
\end{eqnarray}

\noindent
The coefficients $g^{\pm}_m$ are determined after $N$ iterations using:

\begin{equation}
g^{\pm}_m =\sum_{k=1}^N \frac{V^{\pm}_{1 k} V^{\pm}_{m k}}{E^N_L(1^+_k) -E_{g.s.}+E_0}\ .
\label{gm}
\end{equation}

\noindent
In this formula, $V_{m k}$ are the eigenvectors of the tridiagonal Lanczos matrix associated with eigenvalue $E^N_L(1^+_k)$. This approach avoids the need to diagonalize the full many-body Hamiltonian; instead, the energy denominator of Eq.~(\ref{eq1}) is effectively mapped onto the small Lanczos matrix.

The primary computational benefit of the scheme in Eqs.~(\ref{stm})--(\ref{gm}) is that convergence can be tracked iteratively using the Lanczos vectors themselves. Because $g^{\pm}_m$ only requires the diagonalization of a small $N\times N$ matrix, and each iteration only adds a single new overlap to the sums in Eqs.~(\ref{mm}) or (\ref{mp}), storage and processing costs are significantly reduced. This results in an efficiency increase of approximately a factor of two compared to the method in Ref.~\cite{Caurier1990}.

Finally, we acknowledge that shell-model predictions for GT strengths and $2\nu\beta\beta$ NMEs generally require a quenching factor to align with experimental data. This renormalization accounts for many-body corrections to the weak operator, deficiencies in effective interactions, and correlations existing outside the valence space. Our analysis against experimental data~\cite{HoroiStoicaBrown2007,NeacsuHoroi2015,NeacsuHoroi2016,SenkovHoroiBrown2014,Senkov2016} suggests quenching factors between $0.42$ and $0.88$. The optimal values provided in Table~\ref{qfactors} are thus treated as phenomenological calibrations dependent on the specific interaction and valence space.

\section{Results} \label{results}

The strength-function technique described above was tested by comparing running NMEs obtained from the Lanczos expansion with explicit sums over intermediate $1^+$ states. These comparisons are valuable because the running sum is a sensitive diagnostic: individual low-energy and high-energy GT contributions can have different signs, and apparent partial convergence at low excitation energy does not always guarantee the final value. The benchmarks below show that the Lanczos strength-function method reproduces the direct sums where those direct sums can be carried out, while remaining practical in cases where a much larger number of states would be needed.

Table~\ref{qfactors} collects the experimental energy inputs and empirical matrix elements used throughout the calculations. The values of $M^{eff}_{2\nu}$ are those extracted from the recommended experimental half-lives, while $M_{2\nu}$ is obtained from Eq.~(\ref{2nu-hl}) using the free-nucleon value of $g_A$. For nuclei where shell-model calculations are reported below, the last two columns give the effective Hamiltonian and the corresponding optimal GT quenching factor $q_{opt}$ required to reproduce the empirical $M_{2\nu}$. Multiple entries for the same isotope show the interaction dependence of this calibration. Blank entries indicate cases listed for context from the experimental systematics, but not calculated with a specific shell-model Hamiltonian in this work.

The calculations use three standard shell-model valence spaces, chosen according to the mass region. For $^{48}$Ca we use the full $fp$ shell outside a $^{40}$Ca core, with active proton and neutron orbitals $0f_{7/2}$, $0f_{5/2}$, $1p_{3/2}$, and $1p_{1/2}$. The two interactions used in this space are GXPF1A~\cite{Honma2004,Honma2005,HoroiStoicaBrown2007} and KB3G~\cite{KB3G}, which were also used as starting Hamiltonians in the statistical analysis of Ref.~\cite{Horoi-prc22}. For $^{76}$Ge and $^{82}$Se we use the $jj44$ valence space outside a $^{56}$Ni core, with active orbitals $1p_{3/2}$, $1p_{1/2}$, $0f_{5/2}$, and $0g_{9/2}$. The corresponding Hamiltonians are JUN45~\cite{JUN45} and GCN2850~\cite{MenendezPovesCaurier2009}, both based on realistic interactions and subsequently adjusted to spectroscopy in this mass region. For $^{128}$Te, $^{130}$Te, and $^{136}$Xe we use the $jj55$ valence space outside a $^{100}$Sn core, consisting of the $0g_{7/2}$, $1d_{5/2}$, $1d_{3/2}$, $2s_{1/2}$, and $0h_{11/2}$ orbitals. In this space we use the GCN5082~\cite{GCN-int} and SVD~\cite{Chong2012} Hamiltonians. As discussed in Ref.~\cite{NeacsuHoroi2015,NeacsuHoroi2016}, this $jj55$ space omits the spin-orbit partners of the $0g_{7/2}$ and $0h_{11/2}$ orbitals, which is one reason why effective GT quenching remains important.

\begin{figure}
    \centering
    \subfigure[]{\includegraphics[width=0.50\textwidth]{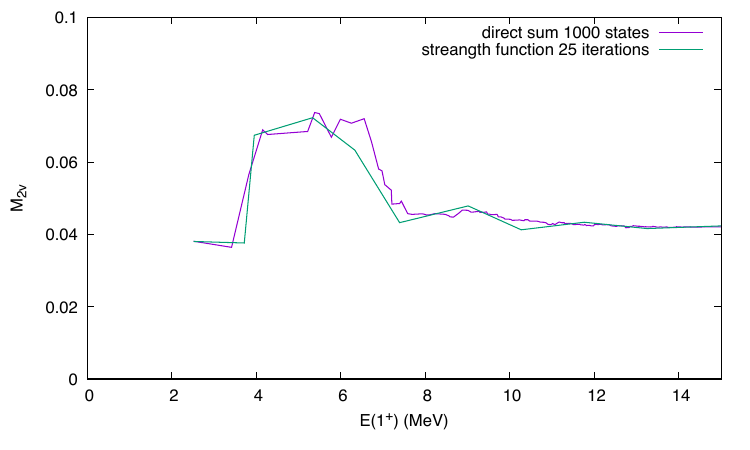}} 
    \subfigure[]{\includegraphics[width=0.50\textwidth]{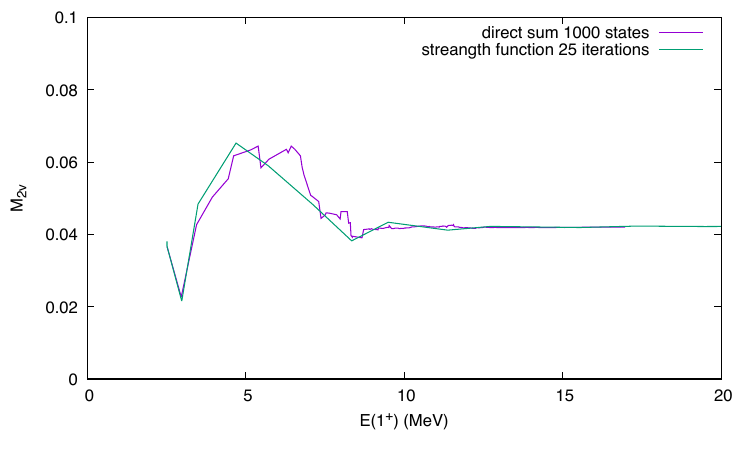}} 
    \caption{Running $M_{2\nu}\left(E(1^+)\right)$ for $^{48}$Ca as a function of the excitation energy of the $1^+$ states in the intermediate nucleus $^{48}$Sc: (a) GXPF1A; (b) KB3G. In each panel the direct sum over 1000 intermediate states is compared with the strength-function result obtained with 25 Lanczos iterations. The plateau at high excitation energy gives the final unquenched shell-model NME used to infer the corresponding quenching factor.}
    \label{ca48}
\end{figure}

Figure~\ref{ca48} shows the benchmark case of $^{48}$Ca. This nucleus is especially useful because the $pf$-shell calculation is small enough that a direct summation over many $1^+$ states in $^{48}$Sc can be performed, while still displaying the cancellations typical of $2\nu\beta\beta$ matrix elements. For both GXPF1A and KB3G, the running NME rises rapidly when the first strong low-lying GT contributions enter, changes again in the region where additional strength appears, and then reaches a stable plateau. The 25-iteration strength-function result follows these features and reproduces the explicit sum over 1000 states. The agreement demonstrates that the relevant information in the GT strength distribution is captured by a relatively small Lanczos space.

\begin{figure}
\includegraphics[width=9cm]{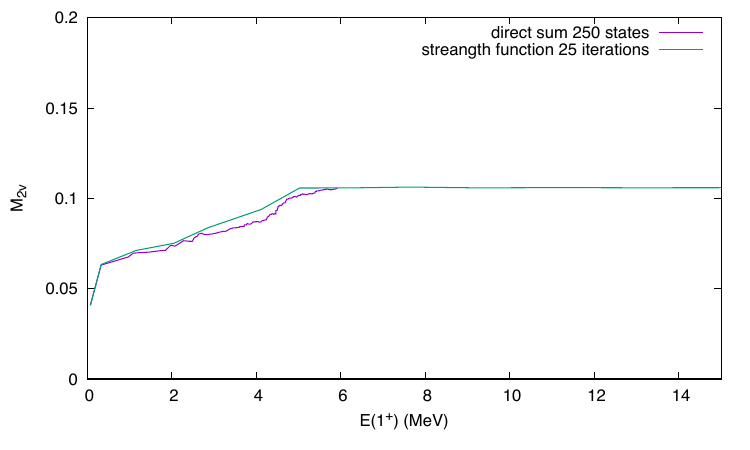}
\caption{Running $M_{2\nu}$ for $^{82}$Se using the GCN2850 effective Hamiltonian. The direct sum over 250 states is compared with the strength-function result obtained with 25 Lanczos iterations. The smooth approach to the plateau indicates that the low-energy part of the $^{82}$Br $1^+$ spectrum dominates the final value, with the higher-energy contribution mainly stabilizing the sum.}
\label{nme-range}
\end{figure}

For heavier nuclei such as $^{82}$Se, the intermediate-nucleus dimensions increase substantially and exhaustive diagonalization becomes less attractive. Figure~\ref{nme-range} compares the strength-function result with a direct sum over 250 states for the GCN2850 Hamiltonian. The direct sum is already close to its plateau after the first few MeV of excitation energy, and the Lanczos calculation reproduces this gradual saturation. This agreement indicates that the method remains stable beyond the light $pf$-shell benchmark and that the dominant low-energy and higher-energy contributions are represented accurately by the Lanczos expansion.

\begin{figure}
    \centering
    \subfigure[]{\includegraphics[width=0.50\textwidth]{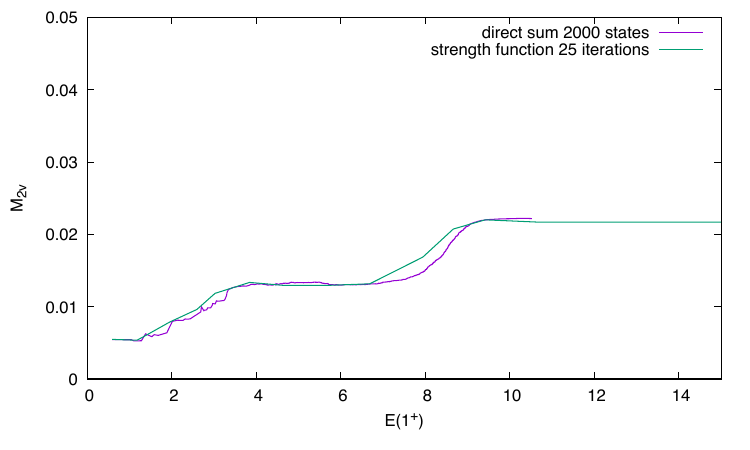}} 
    \subfigure[]{\includegraphics[width=0.50\textwidth]{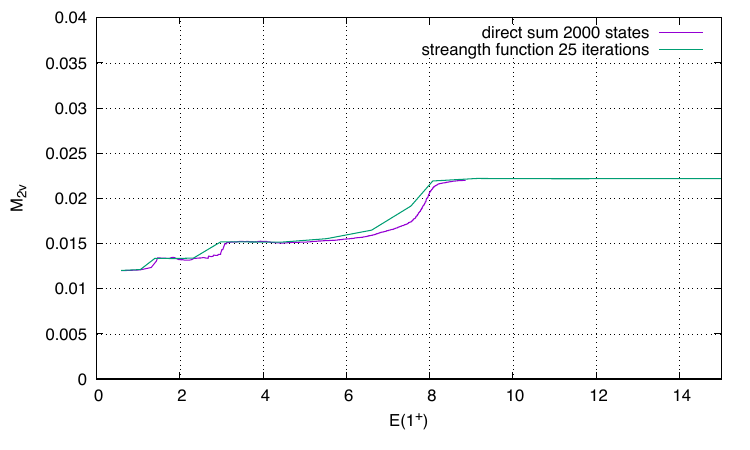}} 
    \caption{Running $M_{2\nu}$ for $^{136}$Xe: (a) SVD effective Hamiltonian; (b) GCN5082 effective Hamiltonian. Both panels compare the direct sum over 2000 $1^+$ states in $^{136}$Cs with the strength-function result obtained with 25 Lanczos iterations. The visible change near the GT giant-resonance region is reproduced by the strength-function method, which is important because the final $^{136}$Xe NME is small and sensitive to cancellations.}
    \label{xe136}
\end{figure}

The method is also applicable to $^{136}$Xe, one of the most important isotopes for current and future experimental programs. Figure~\ref{xe136} compares SVD and GCN5082 calculations with direct sums over 2000 intermediate states. The two Hamiltonians lead to different detailed running patterns, as expected from their different spectroscopy and GT distributions, but the strength-function calculation follows the corresponding direct sum in both cases. This is an important check because the experimentally small $2\nu\beta\beta$ NME of $^{136}$Xe makes the final result particularly sensitive to cancellations.

\begin{figure}
    \centering
    \subfigure[]{\includegraphics[width=0.50\textwidth]{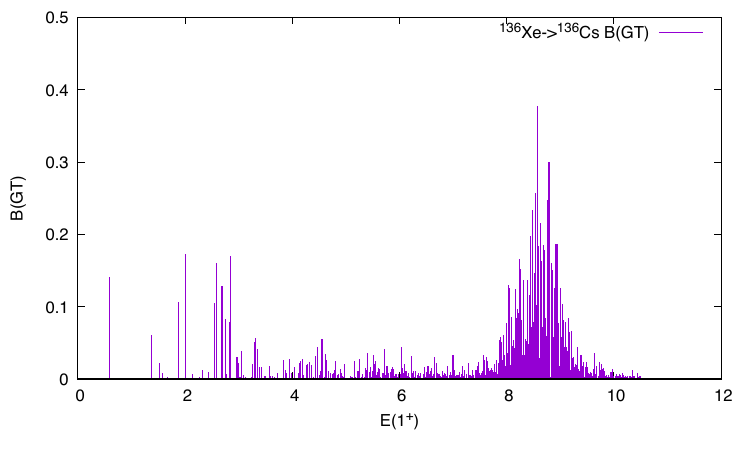}} 
    \subfigure[]{\includegraphics[width=0.50\textwidth]{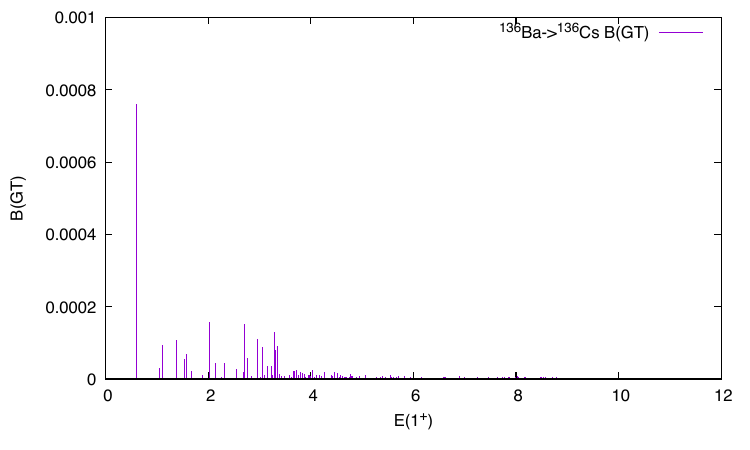}} 
    \caption{GT strength distributions entering the SVD calculation for $^{136}$Xe: (a) $B(GT^-)$ for $^{136}$Xe $\to$ $^{136}$Cs; (b) $B(GT^+)$ for $^{136}$Ba $\to$ $^{136}$Cs. The $2\nu\beta\beta$ NME depends on the product of the two amplitudes through common $1^+$ states, weighted by the energy denominator in Eq.~(\ref{eq1}). The concentration of strength in the giant-resonance region explains the additional structure seen in the running $M_{2\nu}$ of Fig.~\ref{xe136}.}
    \label{ge76}
\end{figure}

Figure~\ref{ge76} displays the two GT strength distributions entering the $^{136}$Xe calculation: the $\beta^-$ branch from $^{136}$Xe to $^{136}$Cs and the $\beta^+$ branch from $^{136}$Ba to $^{136}$Cs. Their overlap through common $1^+$ states determines the numerator of Eq.~(\ref{eq1}), while the energy denominator weights low-lying strength more strongly. The figure also clarifies the origin of the structure seen in the running NME: a visible contribution appears in the region of the GT giant resonance, around several MeV excitation energy, before the sum settles to its final value.

\begin{figure}
    \centering
    \subfigure[]{\includegraphics[width=0.50\textwidth]{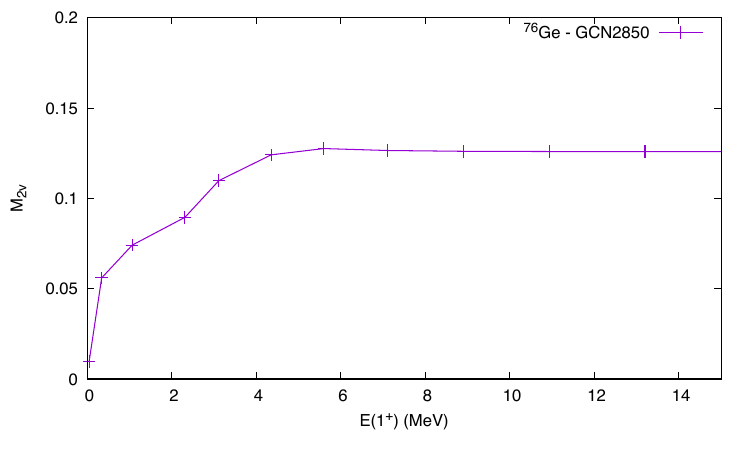}} 
    \subfigure[]{\includegraphics[width=0.50\textwidth]{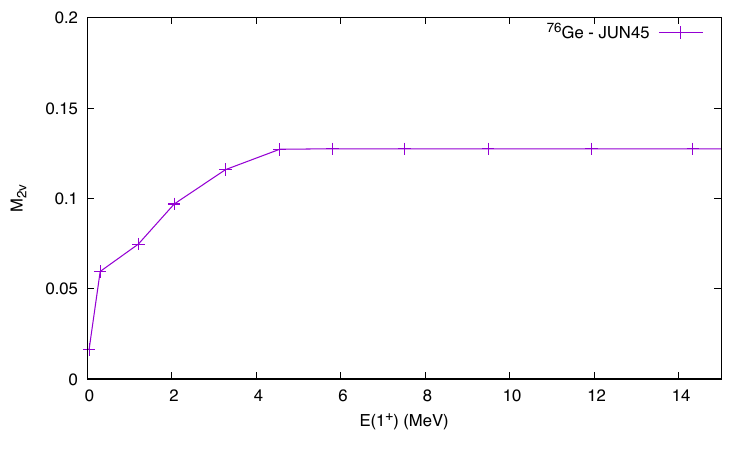}} 
    \caption{Running $M_{2\nu}$ for $^{76}$Ge: (a) GCN2850 effective Hamiltonian; (b) JUN45 effective Hamiltonian. The direct sum over 2000 $1^+$ states is replaced with the strength-function result after 25 Lanczos iterations. The two Hamiltonians give somewhat different running patterns, but in both cases the Lanczos result follows the explicit sum to the final plateau.}
    \label{ge76-res}
\end{figure}

For $^{76}$Ge, another isotope of high experimental interest, we tested both GCN2850 and JUN45. Figure~\ref{ge76-res} shows that the strength-function result converges to the corresponding direct sum for each Hamiltonian. The comparison is useful because the two interactions differ in their detailed description of the low-lying $1^+$ spectrum, yet the numerical method remains robust and does not rely on a special feature of a single Hamiltonian.

\begin{figure}
    \centering 
    \subfigure[]{\includegraphics[width=0.50\textwidth]{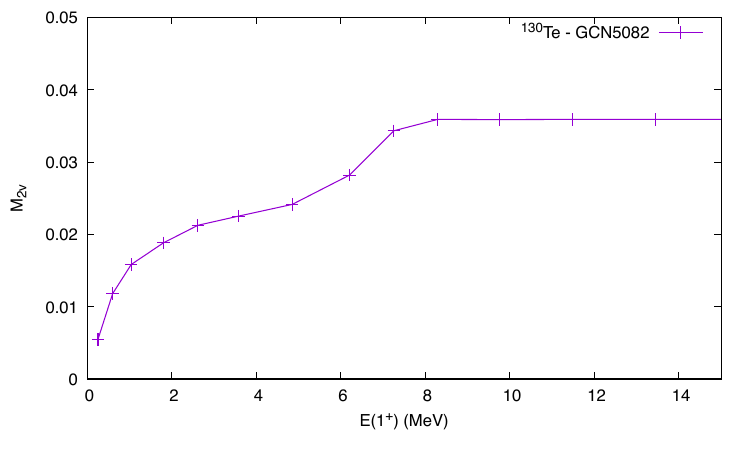}}
    \subfigure[]{\includegraphics[width=0.50\textwidth]{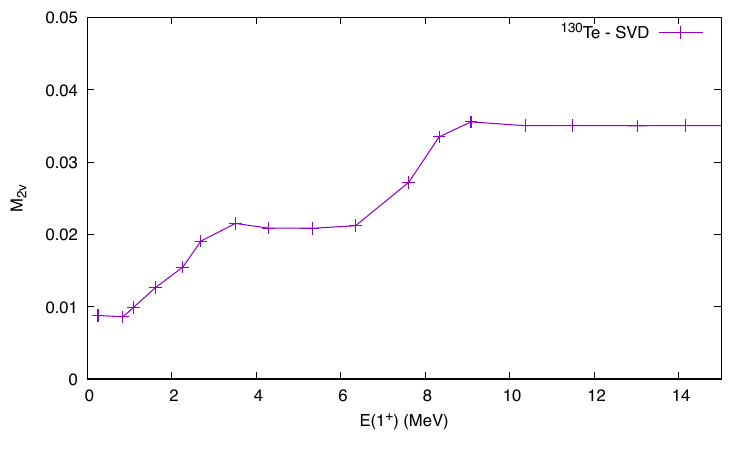}}
    \caption{Running $M_{2\nu}$ for $^{130}$Te: (a) GCN5082 effective Hamiltonian; (b) SVD effective Hamiltonian. These panels test the method in the Te/Xe valence space and show how the final plateau, and therefore the inferred $q_{opt}$ in Table~\ref{qfactors}, depends on the chosen interaction.}
    \label{te130}
\end{figure}

The $^{130}$Te results in Fig.~\ref{te130} provide a similar test in the $50\leq Z,N\leq 82$ mass region. The GCN5082 and SVD calculations again show that the Lanczos strength-function method follows the full running sum and can be used to extract the converged NME without explicitly resolving a large number of intermediate states. The difference between the two final values is reflected in the distinct optimal quenching factors listed in Table~\ref{qfactors}, illustrating that the quenching calibration is not a universal number but depends on the Hamiltonian and model space.

\begin{figure}
    \centering
    \subfigure[]{\includegraphics[width=0.50\textwidth]{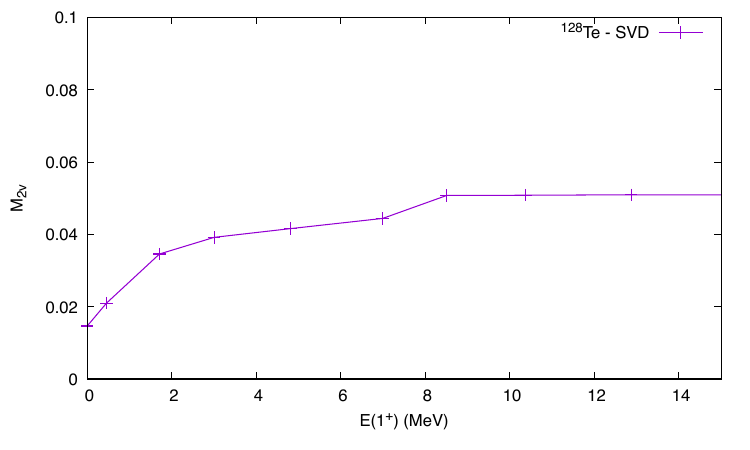}}
    \subfigure[]{\includegraphics[width=0.50\textwidth]{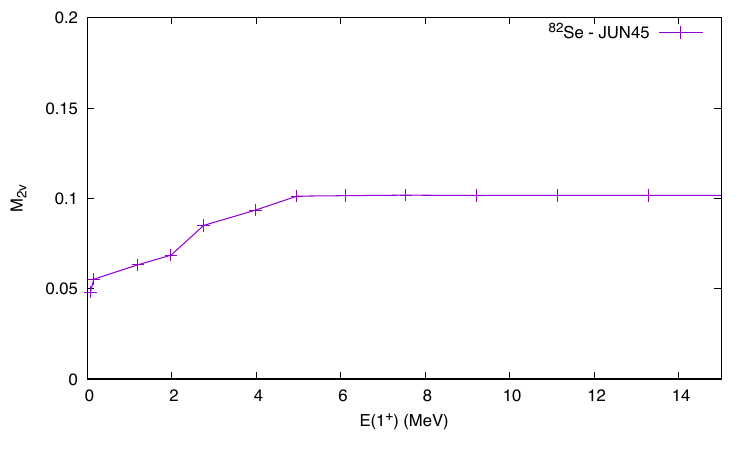}}
    \caption{Additional running-sum tests: (a) $^{128}$Te with the SVD effective Hamiltonian; (b) $^{82}$Se with the JUN45 effective Hamiltonian. These cases complement Figs.~\ref{nme-range} and \ref{te130} by checking the same algorithm for a different isotope in the Te region and a different Hamiltonian in the $A\approx 80$ region.}
    \label{tese}
\end{figure}

Finally, Fig.~\ref{tese} presents $^{128}$Te with SVD and $^{82}$Se with JUN45. Together with the preceding figures, these cases show that the procedure is not tuned to a single isotope or interaction. The $^{128}$Te case is also useful because its small $Q_{\beta\beta}$ value makes the energy denominator comparatively restrictive, while the $^{82}$Se JUN45 result checks that the agreement seen with GCN2850 is not interaction-specific. The same algorithm can be used across the $pf$ shell, the $A\approx 80$ region, and the heavier Te/Xe region, provided that the initial and final shell-model wave functions and the GT doorway vectors are available.

\section{Taylor expansion terms} \label{Taylor}

The standard expression in Eq.~(\ref{2nu-hl}) uses the leading approximation in which the lepton phase space is factorized from the nuclear energy denominator. More refined treatments expand the lepton kinematic factors in powers of the excitation-energy denominator. In Ref.~\cite{Niţescu_2024} we showed that such a Taylor expansion requires additional nuclear matrix elements with higher powers of the same denominator. The first two beyond the leading $M_{2\nu}$ are

\begin{equation}
M_{2\nu-3} = 
\sum_k \frac{\langle 0^+_f\vert q \sigma\tau^-\vert 1^+_{k}
\rangle 
\langle 1^+_{k}\vert q \sigma\tau^-
\vert 0^+_i\rangle}{(E_{rel,1_1^+}(1_k^+) +E(1_1^+)+ E_0)^3} .
\label{eqM3}
\end{equation}
and
\begin{equation}
M_{2\nu-5} = 
\sum_k \frac{\langle 0^+_f\vert q \sigma\tau^-\vert 1^+_{k}
\rangle 
\langle 1^+_{k}\vert q \sigma\tau^-
\vert 0^+_i\rangle}{(E_{rel,1_1^+}(1_k^+) +E(1_1^+)+ E_0)^5} .
\label{eqM5}
\end{equation}

These quantities emphasize low-lying $1^+$ strength even more strongly than $M_{2\nu}$ because the denominators are cubed or raised to the fifth power. They therefore provide a stringent test of the numerical method: small inaccuracies in the first few intermediate states would be amplified relative to the leading NME.

\begin{figure}
    \centering
    \subfigure[]{\includegraphics[width=0.50\textwidth]{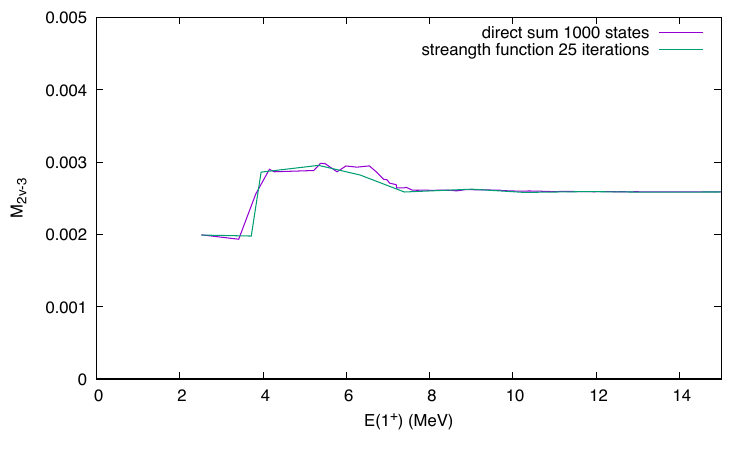}} 
    \subfigure[]{\includegraphics[width=0.50\textwidth]{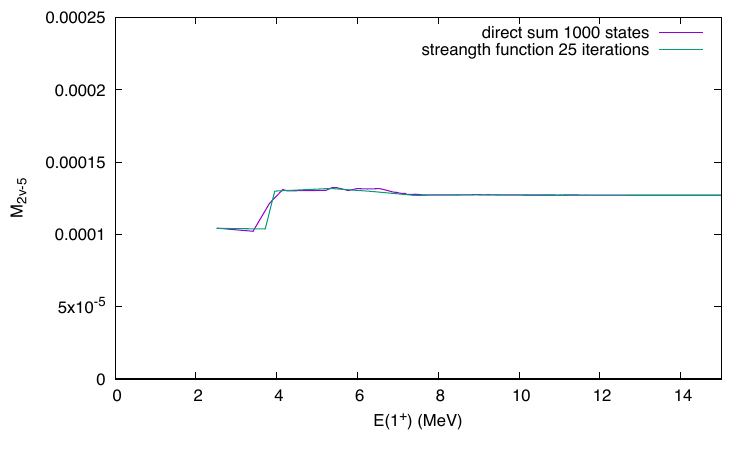}} 
    \caption{Higher-order running NMEs for $^{48}$Ca with the GXPF1A effective Hamiltonian: (a) $M_{2\nu-3}$; (b) $M_{2\nu-5}$. The direct sum over 1000 states is compared with the strength-function result obtained with 25 Lanczos iterations. Because the denominators are raised to the third and fifth powers, these quantities are more sensitive to the lowest $1^+$ states than $M_{2\nu}$, making the agreement a stringent test of the method.}
    \label{M35}
\end{figure}

Figure~\ref{M35} shows the results for $M_{2\nu-3}$ and $M_{2\nu-5}$ in $^{48}$Ca with GXPF1A. The strength-function calculation again agrees very well with the explicit sum over 1000 states. This demonstrates that the same Lanczos coefficients used for the leading NME can be generalized to higher inverse powers of the denominator, making the method applicable to Taylor-expanded $2\nu\beta\beta$ half-lives and to related two-neutrino electron-capture calculations.

The ability to calculate these higher-order terms efficiently is important for precision half-life predictions, especially when the experimental accuracy improves or when the lepton wave functions and atomic effects are treated beyond the leading factorized approximation.

\section{Conclusion and Outlook} \label{conclusions}

We presented an efficient Lanczos strength-function method for calculating $2\nu\beta\beta$ NMEs in the interacting shell model. The method replaces the explicit diagonalization of a large number of $1^+$ intermediate states by a calculation in the much smaller Lanczos space generated from GT doorway states. The required coefficients are obtained from the tridiagonal Lanczos matrix, and convergence can be monitored iteration by iteration using quantities already produced by the algorithm.

Benchmark calculations show that the method reproduces explicit running sums for $^{48}$Ca, $^{76}$Ge, $^{82}$Se, $^{128,130}$Te, and $^{136}$Xe with several effective Hamiltonians. The agreement is obtained even in cases where the running NME displays nontrivial cancellations and contributions from the GT giant-resonance region. The approach therefore provides a practical way to evaluate $M_{2\nu}$ in spaces where direct summation is computationally expensive or impossible.

The same framework was also extended to the higher-order NMEs $M_{2\nu-3}$ and $M_{2\nu-5}$ needed in Taylor-expanded phase-space treatments. The successful $^{48}$Ca benchmark indicates that precision corrections to $2\nu\beta\beta$ and related two-neutrino processes can be treated without introducing a separate computational strategy.

These results are relevant for the broader $0\nu\beta\beta$ program because statistical shell-model studies show a strong correlation between $2\nu\beta\beta$ and $0\nu\beta\beta$ NMEs~\cite{Horoi-prc22}. Accurate $2\nu\beta\beta$ calculations, together with the optimal quenching factors inferred from experiment, can help constrain effective Hamiltonians and guide uncertainty estimates for $0\nu\beta\beta$ NMEs. The efficiency of the present method should also make it possible to explore enlarged model spaces, such as calculations including $2\hbar\omega$ excitations in the $sd$-$pf$ region, where direct diagonalization of the intermediate nucleus is not feasible.

Future work will focus on applying the method in such enlarged spaces and on comparing the resulting correlations with those found in emerging ab initio calculations~\cite{Yao2020,Novario2021,Belley2021}. As ab initio methods continue to improve, strength-function shell-model calculations can provide complementary large-scale benchmarks, broad statistical samples, and phenomenological constraints from measured $2\nu\beta\beta$ half-lives.

\vspace{0.5cm}
    {\it Acknowledgements}. The author acknowledges support from the US Department of Energy grant DE-SC0022538 "Nuclear Astrophysics and Fundamental Symmetries". 

\bibliographystyle{apsrev4-2}
\bibliography{bb-n}

\end{document}